# Smart Support for Mission Success

Juliette MATTIOLI[1][0000-0001-9775-2716] and Pierre-Olivier ROBIC[2]

[1] Thales, France
[2] Thales Global Services, France
{juliette.mattioli,pierre-olivier.robic}@thalesgroup.com

**Abstract.** Today's battlefield environment is complex, dynamic and uncertain, and requires efficient support to ensure mission success. This relies on a proper support strategy to provide supported equipment able to fulfill the mission. In the context of defense where both systems and organization are complex, having a holistic approach is challenging by nature, forces and support agencies need to rely on an efficient decision support system. Logistics, readiness and sustainability are critical factors for asset management, which can benefit from AI to reach "Smart In Service" level relying especially on predictive & prescriptive approaches and on effective management of operational resources. Smart Support capacities can be then monitored by appropriate metrics and improved by multi-criteria decision support and knowledge management system. Depending on the operational context in terms of information and the objective, different AI paradigms (data-driven AI, knowledge-based AI) are suitable even a combination through hybrid AI.

## 1 Smart Support Capacities

Availability and reliability are key if military forces ensure mission success and lifecycle cost mastery by enhancing force readiness and support chain. To achieve this, military forces and support agencies are increasingly having to battle to stay on top on three fronts – support services, performance and maintenance – to maintain their effectiveness. The aim of Smart Support is to last longer and to require less support, thereby reducing costs and increasing return on investments. The impact is assessed in terms of key performance indicators (KPI) such as reliability, availability, maintainability and testability (RAMT), Life Cycle Costs and sometimes System Safety (RAMS).

The main objective of this paper is to underline how artificial intelligence (AI) could enhance these aspects of supportability throughout the operational life cycle of defense equipment, becoming the difference between mission success and failure (see **Fig. 1**) by:

- **Optimizing Operational and Financial flow**: by **monitoring asset** availability rates and link policy changes to maintenance outcomes, AI-based solutions achieve significant reductions sustainment costs. Moreover, an efficient **asset management** [1] and supply chain can identify the relevant resources and anticipate maintenance through in-theater visibility of operational, intermediate and depot maintenance and real-time total asset visibility of globally deployed assets.

- **Managing the Performance/Delay/Cost Balance**: In order to monitor SLA (Service Level Agreement), **stakeholder's satisfaction** and improved performances balance, accurate and relevant **KPIs** are mandatory. Moreover, **knowledge management** and know-how capitalization are key for diagnostic and repairs toward automated prescription and to improve transfer of experience in the context of turnover.

- **Minimizing unavailability**: Effective plan and schedule maintenance have to be based on real-time mission and asset utilization data, resource constraints and parts availability, without incurring costs due to oversupply. It covers the following maintenance activities [2][2]: (1.a) **Reactive maintenance** (a.k.a Curative) is performed when equipment has failed. One issue is to find correlations to support the diagnosis of a failure or to build a root-cause analysis. (1.b) **Preventive Maintenance**, aims to reduce the probability of failure on equipment through regular overhaul. (2) **Predictive Maintenance** is performed before equipment failure using predictive insights. To compute the Remaining Useful Life (RUL) of equipment is a cornerstone of such maintenance. (3) **Prescriptive Maintenance** occurs once the failure is predicted, solutions are provided to identify what action to take to improve the outcome. It answers questions such as "what is the best course of action and which risks will it involve?". Moreover, applying root cause analysis will also eliminate recurring sources of quality defects.

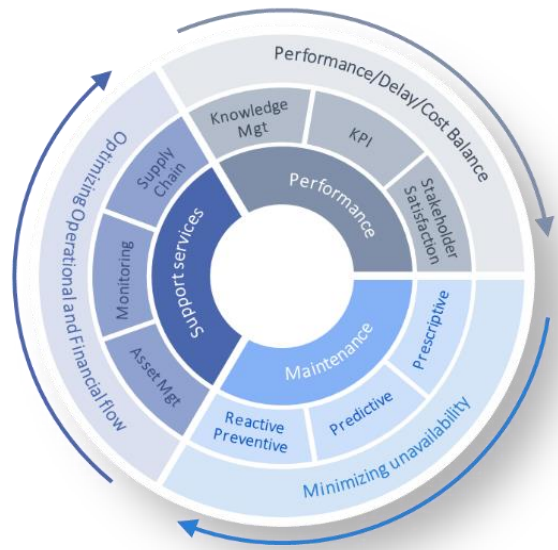

**Fig. 1.** Main defense Support capabilities

In the following, we distinguish between three major paradigms in AI:

- **Knowledge-based AI** is the branch of AI that simulates the mind's symbolic processing by attempting to explicitly represent human knowledge in a declarative form (i.e. using facts and rules). Thus, to be successful in producing human-like intelligence, it is necessary to translate implicit or procedural knowledge into an explicit form using symbols and rules for their manipulation. Therefore, symbolic approaches are based upon a syntax that is endowed with formal semantics (meaning) useful for properties expression and verification. They have been successfully applied to increase the system's reliability in different application domains such as prescriptive maintenance [4].
- Contrary to Knowledge-based AI, **data-driven AI** is not built upon an explicit representation of human expertise: the behavior is learnt from examples/data. Usually, we distinguish between two categories of problems that can be addressed:
  - **Classification problem**: find a function which maps an input to a discrete output, also called label or class. This problem is referred to as binary classification when there are two classes, and multi-class classification when there are more than two classes. Image recognition is a typical example of multi-class detection, when a user is interested in getting what is represented in an image (e.g., default recognition).
  - **Regression problem**: find a function which maps an input to a continuous output. A typical example is to "reproduce" an unknown function when given a large collection of inputs and corresponding outputs (e.g., RUL estimation).

In the data-driven AI paradigm, connectionist AI algorithms are based upon a statistical or probabilistic model which is tightly coupled to data sets which are first used for model training, and once tuned, for performance evaluation. The model is often structured as a set of nodes defined by multi-value functions or random variables. The nodes are interconnected, and the links can be randomly weighted by values influencing nodes inputs/outputs [2]. On one side, data-driven AI approaches successfully characterize and capture the salient traits of the data sets. However, being the connectionist models heuristic and agnostic of typical notion-encapsulation archetypes, they lack argumentation necessary for explainability. On the other side, symbolic approaches introduce a semantic layer aligned with the notion-encapsulation archetype, which is amenable for expressing domain knowledge and concerns useful for validation and argumentation. More recently, **hybrid AI** approaches integrating knowledge-based and connectionist paradigms have been proposed. Hybrid approaches aim to profit of salient features of both symbolic and connectionist leaving out any potential concurrence [5]. In certain cases, the complementary between techniques even leads to overcome certain limitations of each other.

## 2 AI-based asset management

Based on ISO 55000, asset management (AM) is described as coordinated activity of an organization to realize value from assets. AM is one of the ways to improve the management strategy for different the phases of asset life cycle, in order to optimize its lifetime, to reduce costs, to improve supply chain management, to extend health of assets, to enhance safety and to decrease unplanned downtime. Military operations are renowned for precision, focus, efficiency and accuracy. The only way to achieve such critical issues is to deploy real-time military AM, tracking and maintenance solution to provide spare and repairs, and more generally equipment, in timeliness into a dynamic and uncertain context. Moreover, it is difficult to design such asset management support that can handle various operational risks, which can be responsible for significant losses and damage. Thus, to address asset management as a whole, various AI technologies are needed as illustrated in **Fig. 2**.

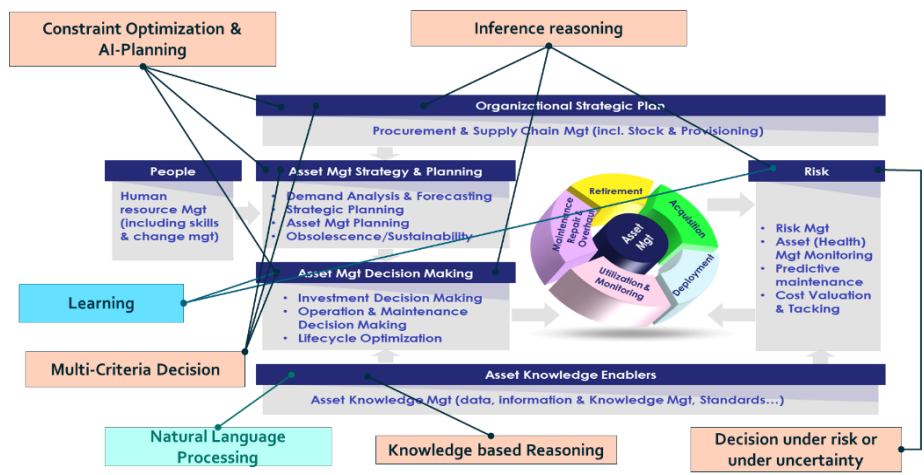

**Fig. 2.** AI-based Asset Management (from [1]).

Fuzzy Inference methods [6] can enhance decision-making processes by quantifying the severity of a risk and assists stakeholders in taking further actions that mitigate this severity level. It will support through planning and monitoring acquisition, use, care and/or retirement of assets through related risks and costs mitigation over their overall life assisting in increased reliability and accountability.

Moreover, effectively managing the supply chain is challenging for optimizing the supply chain flows in their planning and scheduling through upstream and downstream linkages and addressing impacts of uncertainties in component supply. Designing such advanced planning system requires specific skills in knowledge-based AI methods such as constraint programming (CP). Recall that planning activity determines the sequence of tasks (the plan), the scheduling activity decides on the specific resources, operations and their timing to perform the tasks (the schedule) and the resource management activity controls at specified time the resource allocation. Nowadays, CP is mostly applied to resolve combinatorial and optimization problems [7] that need a real-time solution as planning and scheduling, sequencing, and configuration problems. Here, resources may be raw materials, engines or engine components, people or any other object that is able to perform a useful operation and tasks represent steps in the overall asset management process. An allocation plan is represented by a set of points illustrating which resources are assigned to each task. However, all the points of the task / resource plan cannot be feasibly assigned. The problem's constraints indicate which points are feasible and which are to be discarded. In this way, to solve a resource allocation problem, you search for a set of points within the task/resource plan, which satisfy both the problem constraints and the objectives. CP is a key approach to define tactical support plan against dynamic constraints in terms of resources.

Modeling and predicting asset RUL becomes a particularly challenging exercise when derivatives are involved. As a result, constructing optimal spare part portfolio for an equipment that include derivatives is difficult, because their RUL are not well defined and are contingent on other assets. Most forecasting approaches rely historical data. This is a realm where data-driven AI can play a role. For example, machine learning such as deep learning can be used.

## 3 Knowledge engineering for smart support capacities

In its initial form, Knowledge Engineering (KE) focused on the transfer process; transferring the expertise of a problem-solving human into a program that could take the same data and make the same conclusions. Thus, KE dealt with the development of information systems in which knowledge and reasoning play pivotal roles. In the 1990s, the attention of the KE community shifted gradually to domain knowledge, in particular reusable representations in the form of ontologies. This evolution aimed at alleviating KE limitation to accurately reflect how humans make decisions and more specifically its failure to take into account intuition known as "reasoning by analogy". Today, KE refers to the process of understanding and then representing human knowledge in data structures, semantic models (conceptual diagram of the data as it relates to the real world), and heuristics (rules in AI context). KE goal it to acquire the knowledge related to a specific task from expert people, and to transform it into if-then rules in order to make deductions or choices. The basic assumption is that both knowledge and experience can be captured and archived in textual or rule-based form, using formalization methods.

When maintainers do not have experience on the equipment's behavior, they typically revert to the equipment vendor for advice. The latter tends to be very conservative as it has to provide one single shutdown frequency that covers all various types of operations and conditions. Furthermore, without specific knowledge about the asset condition, maintainers will have the tendency of over-maintain an asset with the objective of reducing potential unplanned failures. Another challenge is induced by competencies attrition for through life-cycle of defense equipment for long term support. Thus, one could envision knowledge engineering providing easy access to the various domains of knowledge required. Being aware of environments in the field will enable equipment monitoring in more effective ways, making life easier for the operators.

For example, [8] uses knowledge of equipment's life-cycle loading conditions, mechanical or thermal features within the contextual usage in Prognostic Health Monitoring. Then, the knowledge of failure causes or of the process lead to the occurrence of failure allows us to have an accurate RUL estimation. In [9], an adaptive network-based fuzzy inference system is used to obtain Overall Equipment Effectiveness measure through information on availability, performance efficiency and rate of quality.

To reach this goal, one must however extract how an expert reasons and decides while performing his/her usual tasks. As such, knowledge extraction can be achieved through several ways. First, interviews can be conducted among expert to gather explanations of their decision-making processes and to model business rules (knowledge-based system). Knowledge based approach and more precisely an ontology-based approach proposed in [10] is used not only to automatically capture the relations between the entities in the domain context knowledge characterizing the failure but also to capture the information and establish relations from maintenance reports. All information has been encoded in a knowledge graph for encoding its semantics, and then the problem is addressed as a link prediction problem in a knowledge graph to support prescriptive maintenance. The approach proposed in [10] differs in that it can use historical fault description to proposing solutions for new unseen fault description.

However, such knowledge-based AI approaches could be imprecise or incomplete due to the implicit and subjective nature of some knowledge. An alternative relies on digital twin for knowledge capturing based on machine learning techniques. Recently, Thales developed the "Cognitive Shadow©" system [11], a prototype tool that automatically learns a user's decision pattern from past decisions and that can provide advisory warnings in real time when the decision of the user does not match the system prediction.

## 4 Multi-criteria decision to design and assess key mission success drivers

Most studies on specific military missions such as humanitarian aid and disaster relief missions suggest that the quality of mission preparation [12] and service support (including logistics and asset management) will affect their success [13]. Thus, to optimize the performance/delay/cost tradeoff, stakeholders face three main questions [14]: How to build relevant mission success indicators (performance measurement) fulfilling the criteria of cost, efficiency, reliability, availability, maintainability, safety...; how to ensure that high-quality multi-criteria indicators are produced? How to exploit such indicators for improving military service support capacities? To support such issues, we follow an approach based on a preference model to elicit the criteria and metrics called MYRIAD© composed of the following stages:

- **Structuring phase** aiming to represent a hierarchy of concerns by a tree in which the root represents the overall evaluation and the leaves are the elementary metrics such as Overall Equipment Effectiveness (OEE), cost of maintenance and repair, productive value per equipment. All nodes except the leaves return a numerical evaluation that is a satisfaction degree.
- **Criteria construction** consists in quantifying the evaluation tree on the criteria nodes. Judgment for each metric is quantified separately by a utility function. This has to be aligned with Operational concepts (CONOPS, CONEMP & CONUSE) & mission specificities.

- **Aggregation construction** consists in quantifying the evaluation tree on each aggregation node to assess the overall support key drivers. One needs to aggregate the partial evaluations to obtain higher-level evaluations. Most usual aggregation function used is the weighted sum. Its main drawback is that it fails to represent real-life decision strategies including veto or interaction among criteria. We use a model able to represent veto, favor, complementary among criteria and so on [15]. At the end of this stage, the preference model is thoroughly elicited.

The preference model is applied on derived models (variants) that stakeholders wish to assess and compare. They are not only interested in the evaluations, but also in the explanations of these scores [16]. Above Myriad©, e-Myriad© is an application which provides insights into a particular evaluation outcome, to explain in an intelligible way, the estimated evaluation score and on which attribute worked to increase the ROI, even if the mechanisms of aggregation are based on notions or concepts which escape the understanding of a non-mathematician. An instance of multi-criteria decision model, the Choquet integral is widely used in real-world applications, due to its ability to capture interactions between criteria while retaining interpretability. This part is in line with the concept of "*Explainable AI*" (a.k.a. XAI) in which AI and how it comes to its decisions are made intelligible to the user by providing explanations [17]. Aimed at a better scalability and modularity, [18] presents a machine learning-based approach for the automatic identification of hierarchical multi-criteria decision models, composed of 2-additive Choquet integral aggregators and of marginal utility functions on the raw features from data reflecting expert preferences.

## 5    Conclusion

As support operation is subject to a wide area of activities and face uncertainty of failure, the monitoring of the performance is by essence quite complex. In order to provide sound capacities, we have to address in a holistic way this overall complexity. This lead to a combination of many AI technics based on data-driven and knowledge-based approaches in order to handle the overall situation and be relevant for purpose. This approach should be deeply useful in a mission portfolio management perspective relying on a fleet of systems as the mission success rely on both the supported equipment and the operation system (socio-technical system / organization operating the system). This is even reinforced when these systems can be subject to dynamic allocation in terms of equipment / capacities. An example of technics is cannibalization where a component of an asset is utilized to provide spare parts for another, is commonly utilized to maintain readiness when spares are not available or scarce.